\documentclass[12pt]{article}
\pdfoutput=1

\usepackage{color}

\usepackage{comment}
\usepackage{graphics}
\usepackage[dvips]{graphicx}
\usepackage{mathrsfs}
\usepackage{amssymb}
\usepackage{amsmath}
\usepackage{verbatim}
\usepackage{float}
\usepackage{slashed}
\newcommand{\be}{\begin{equation}}
\newcommand{\ee}{\end{equation}}
\newcommand{\bea}{\begin{eqnarray}}
\newcommand{\eea}{\end{eqnarray}}

\def\4vol{{\int d^4x \sqrt{-g}}}

\def\beq{\begin{equation}}
\def\eeq{\end{equation}}
\def\bea{\begin{eqnarray}}
\def\eea{\end{eqnarray}}
\def\bitem{\begin{itemize}}
\def\eitem{\end{itemize}}

\newcommand{\nc}{\newcommand}
\nc{\nt}{\tilde{N}}
\nc{\ra}{\rightarrow}
\nc{\lsim}{\begin{array}{c}\,\sim\vspace{-21pt}\\< \end{array}}
\nc{\gsim}{\begin{array}{c}\sim\vspace{-21pt}\\> \end{array}}
\nc{\tnt}{\tilde{N}}
\nc{\tst}{\tilde{t}}

\nc{\LL}{L}
\nc{\vv}{\tilde{v}}

\title{
\vspace*{5mm} \Large\textbf{Magnetic Fluffy Dark Matter}
\vspace*{1.0cm}
\author{\textbf{Kunal Kumar$^a$, Arjun~Menon$^{b,c}$, Tim M.P. Tait$^{d,e}$
}\\\\
\normalsize\emph{$^a$Department of Physics and Astronomy, Northwestern 
University,Evanston, IL 60208}\\
\normalsize\emph{$^b$Physics Division, Illinois Institute of Technology, Chicago, IL 60616, USA}\\
\normalsize\emph{$^c$Institute of Theoretical Sciences, University of Oregon, Eugene, OR97401, USA}\\
\normalsize\emph{$^d$Department of Physics and Astronomy, University of 
California, Irvine, CA 92697}\\
\normalsize\emph{$^e$Kavli Institute for Theoretical Physics, University of California, Santa Barbara, CA 93106, USA}
}}                
\date{\today}
\begin{document}
\setcounter{page}{0}

\maketitle
\begin{abstract}
We explore extensions of inelastic Dark Matter and Magnetic inelastic Dark 
Matter where the WIMP can scatter to a tower of heavier states.
We assume a WIMP mass $m_\chi \sim \mathcal{O}(1-100)$~GeV and a constant 
splitting between successive states $\delta \sim\mathcal{O}(1 - 100)$~keV. 
For the spin-independent scattering scenario we find that the direct experiments
CDMS and XENON strongly constrain most of the DAMA/LIBRA preferred parameter 
space, while for WIMPs that interact with nuclei via their 
magnetic moment a region of parameter space corresponding to $m_{\chi}\sim 11
$~GeV and $\delta < 15$~keV is allowed by all the present direct detection 
constraints.
\end{abstract}

\thispagestyle{empty}
\newpage
\setcounter{page}{1}

\section{Introduction}
\label{sec:Intro}

The nature of dark matter is one of the fundamental questions facing 
physics, with numerous experiments being performed to 
search for it both directly and indirectly. 
Direct detection experiments hope to detect 
Weakly Interacting Massive Particles (WIMPs) by observing the recoil of target nuclei 
which interact with ambient WIMPs in the nearby galactic halo. 
One particular approach boils down to a counting experiment, looking for a signal in
excess of all known background processes.
Due to the expected small scattering cross-sections involved,
these experiments typically need to reject many large backgrounds 
such as natural radioactivity and cosmic rays in order to be sensitive to dark matter scattering. 

Another approach to direct detection of dark matter relies on the relative motion
of the Earth through the dark matter halo. As the 
Earth orbits the Sun, which in turn is moving about the center of 
the Milky Way, the flux of WIMPs impinging on the Earth undergoes an annual 
modulation~\cite{Drukier:1986tm}.  In particular, it is more likely to find WIMPs at high
relative speeds when the Earth's motion is aligned with the Sun's (in the summer) than when
they are pointing in opposite directions.  A larger relative speed results in an increased potential for
higher energy nuclear recoils.  Given the finite energy thresholds of direct detection experiments,
the result translates directly into an annual modulation of the rate of WIMP scattering with target nuclei.
This modulation helps isolate a WIMP signal from the known backgrounds, which are
not expected to display a strong modulation, and allows for detection of WIMPs without having to identify 
individual events as arising from signal or background. 
In fact, the DAMA/LIBRA experiment has reported evidence for just such 
an annual modulation signal whose peak is in June, consistent with the expectations of dark 
matter scattering~\cite{Bernabei:2008yi}. However, in the simplest dark matter models the typical 
cross-sections needed to generate a modulation signal large enough so as to explain the DAMA/LIBRA results
are so large that they are inconsistent
with the null results at other direct detection experiments, including
XENON100~\cite{Aprile:2011hi} and CDMS II~\cite{Ahmed:2009zw}. 

Inelastic Dark Matter (iDM)~\cite{TuckerSmith:2001hy,Han:1997wn} models attempt to resolve
this puzzle. iDM models alleviate direct detection experimental constraints by 
requiring inelastic scattering of the form 
\bea
\chi + N \to \chi^* + N
\eea 
where $\chi$ is the incoming WIMP, $\chi^*$ is a heavier state into which it must scatter when interacting
with a target nucleus $N$.  Only scattering which transfers at least energy 
$\delta = m_{\chi*} - m_\chi$ is kinematically allowed, which gives more relative weight to large velocity scattering,
enhancing the annual modulation effect.  In addition, since the momentum transfer of the scattering is controlled in
part by the target mass (assuming it is not much greater than $m_\chi$), 
the scattering rate is also different for different mass target nuclei.
Initially, for $m_\chi \sim 100$~GeV and $\delta \sim 100$~keV, iDM models were able to explain the DAMA/Libra
observations, while not being ruled out by other experiments.  In the time since they were proposed, the increases in
sensitivity have since closed the window of parameter space which could explain DAMA.  For example,
XENON100 has ruled out the iDM parameter space with a $\delta \lsim 120$~keV that explains
DAMA as scattering off of Iodine at the 90$\%$ confidence level~\cite{Aprile:2011ts,Alves:2010pt,Farina:2011bh}.

In addition to its atomic mass number, another feature which distinguishes target nuclei are their electromagnetic
properties, including charge and magnetic moment.  Magnetic inelastic Dark Matter (MiDM)~\cite{Chang:2010en} 
exploits these differences by positing an inelastic WIMP whose primary interaction portal with Standard Model (SM)
particles is by exchanging photons by virtue of a magnetic dipole moment
\cite{Bagnasco:1993st,Sigurdson:2004zp,Barger:2010gv,Masso:2009mu,Banks:2010eh}.  
Such a photonic portal favors target nuclei
with larger charge and/or magnetic moment, and leads to an enhanced rate at DAMA, as compared to 
CDMS (for example) since Sodium and Iodine have high magnetic dipole moments in comparison 
to Germanium (see Table~\ref{tab: isotopes}).  A viable MiDM parameter space results
\cite{Chang:2010en,Lin:2010sb}, subject to mild (and somewhat model-dependent) constraints
from the null results of the Fermi/GLAST search for gamma-ray lines \cite{Abdo:2010nc}
from WIMP annihilation \cite{Goodman:2010qn}
and from LEP searches \cite{Achard:2003tx} for missing momentum \cite{Sigurdson:2004zp,Fortin:2011hv,Fox:2011fx}.

Aesthetically, the need for a small splitting $\delta$ relative to other mass scales in the theory such as $m_\chi$
is somewhat mysterious.  The existence of such a
splitting may be motivated by introducing a weakly broken symmetry 
\cite{Chang:2010en,ArkaniHamed:2008qn,Pospelov:2008qx}
which would otherwise require the elements of a WIMP multiplet to be degenerate,
or can be produced in models where the WIMP is a composite state, bound
either by a confined \cite{Alves:2009nf,Kribs:2009fy}
or by a weak long range force \cite{Kaplan:2009de}.  Such models naturally accommodate multiple mass scales, but it
remains true that one tunes a parameter in order to generate a fine or hyper-fine splitting of the correct
size to generate a realistic iDM model.

In this article, we explore a variation of composite inelastic models with a new feature ameliorating the need to
tune any parameter related to $\delta$.  We consider a class of models in which the WIMP is a bound state
of a new confined gauge force, where the dark matter is the lowest
lying state consisting of a heavy preon which provides the bulk of the WIMP mass (and perhaps ``flavor" quantum
numbers which insure its stability), as well as some other
light fundamental degrees of freedom.  The confinement scale $\Lambda$ of the new gauge sector satisfies
$\Lambda \ll m_\chi$, such that one can expect a continuum of excited states whose levels are much smaller
than the WIMP mass itself.  $\Lambda$ is also much smaller than the typical scale required of iDM in order to explain the
DAMA results.  The collective up-scattering from ground state into a variety of the excited states will explain
the DAMA modulation results, with the characteristic splitting scale emerging organically from a theory whose
characteristic splittings are somewhat smaller.  Since this dark matter candidate looks something like a massive
preon surrounded by a rather flimsy (easily perturbed) gluosphere, we refer to it as ``fluffy" dark matter (fDM).
 
We assess the conditions under which fluffy dark matter can fit the DAMA modulation signal by
interacting with nucleons either
through a (canonical) Higgs-like portal, or by virtue of a magnetic moment through the
photon portal.  We model the fluffy spectrum as consisting of a ground state with a continuum of excited states at
roughly evenly spaced intervals above it.   
As we will see shortly, the upshot is that the Higgs-like portal is ruled out by a
combination of XENON100 \cite{Aprile:2011ts} 
and low threshold CDMS \cite{Ahmed:2010wy} data, but magnetic fluffy dark matter (MfDM) is viable
for WIMP masses around $m_\chi \sim 11$~GeV and $\delta \sim$~keV.
 
\section{Effective Theories of Fluffy Dark Matter}
\label{sec:setup}

Fluffy dark matter can arise as the low energy limit of a variety of UV theories.  While it would
be interesting to pursue some detailed examples (allowing one to discuss collider signals, which
could show features of unparticles \cite{Georgi:2007ek} or a hidden valley models \cite{Strassler:2006im}), we defer
construction of detailed models to future work.  Instead, 
we focus on the low energy
dynamics, which we express in terms of an effective field theory containing the composite bound states.  

As an underlying picture, we imagine a large $N$ confining gauge theory, with confinement scale $\Lambda$.  We refer to the vector bosons associated
with the new gauge sector as ``gluons" (without confusion with respect to
the force carriers of the $SU(3)_c$ of the SM).
There are also matter fields consisting of a heavy
adjoint Majorana fermion (the ``gluino")
which is a singlet under the SM gauge interactions, as well as
some connector fields with both SM gauge quantum numbers, and transforming under the new force such
that they can mediate interactions between bound states of the new gauge force and either the SM Higgs
or photon by inducing a magnetic moment.  It would be interesting to explore candidate models in terms of
RS-like dual theories \cite{ArkaniHamed:2000ds}, 
but we leave such constructions for future work.

The lowest lying states of the confined sector consist of glueballs with masses of order $\Lambda$ (which typically can decay into SM particles) and 
glueballinos, $\chi_i$, Majorana fermion bound states of
gluinos and gluons.  We assume an underlying flavor symmetry which renders
the gluino, and thus the $\chi_i$, stable.  
The WIMP is identified as
the lightest of these states with mass $m_\chi$, and there is a sector of excited states with masses
larger by order $\Lambda$.  We will assume that the spectrum of excited states consists of a tower of states
labelled by an integers $n$ with masses,
\bea
m_n = m_\chi + n~ \delta
\eea
where $\delta \sim \Lambda \ll m_\chi$.

The connector fields are responsible for inducing interactions between the bound state WIMPs and the SM
fields.  One such interaction is a scalar coupling to two Higgs doublets,
\bea
r_\chi ~ \overline{\chi}_i \chi_j \, H^\dagger H ,
\eea
where $H$ is the SM Higgs.  After electroweak breaking, this leads to an interaction of $\overline{\chi}_i \chi_j$ with
a single Higgs boson with strength $r_\chi v$.  We have simplified the discussion by assuming that a single
parameter $r_\chi$ controls the interaction, independent of $i$ and $j$.  In principle, this interaction can
mediate elastic as well as inelastic scattering, but since we will find this case has difficulty explaining the DAMA
results in the light of other null searches anyway, we will not dwell on this point here and just assume that
something forbids the $i = j$ terms from occurring.  We refer to a model coupled in this way as fluffy dark matter (fDM).

The second interaction we consider is a magnetic moment interaction,
\bea
\mu_\chi ~\overline{\chi}_i \sigma_{\mu \nu} \chi_j \, F^{\mu \nu} ,
\eea
where $\mu_\chi$ parameterizes the strength of the magnetic dipole, 
$\sigma_{\mu \nu} \equiv i [ \gamma_\mu, \gamma_\nu ] /2$, and $F^{\mu \nu}$ is the photon field strength.
For a Majorana fermion, this interaction vanishes for $i = j$, which nicely explains why the WIMP should
scatter inelastically off of nuclei.  We have again made the simplifying 
assumption that the same parameter $\mu_\chi$ at least approximately
describes the strength of the interaction for all $i$ and $j$.  We refer to a fluffy WIMP which interacts
with the SM primarily in this way as magnetic fluffy dark matter (MfDM).

\section{Direct Detection of Fluffy Dark Matter}

\subsection{Dark Fluffy Scattering}

\subsubsection{fDM}

For an inelastic scattering, the differential scattering rate may be
written \cite{Finkbeiner:2009ug},
\bea
\frac{dR}{dE_R d \cos \gamma} = \frac{\kappa F^2(E_R)}{n(v_0,v_{\rm esc})}
\pi v_0^2 \left[\exp \left(-\frac{(\vec{v}_E \cdot \hat{v}_R + v_{\rm min})
}{v_0^2} \right) - \exp \frac{v_{\rm esc}^2}{v_0^2} \right]\Theta(v_{\rm esc} - 
|\vec{v}_E \cdot \hat{v}_R + v_{\rm min}|) ~~
\label{dRdErdcosgamma}
\eea
where,
\bea
\kappa = N_T \frac{\rho_\chi}{m_\chi} \frac{\sigma_n m_N}{2\mu_n} \frac{(f_p Z+
(A-Z)f_n)^2}{f_n^2},
\eea
and $N_T$ is the number of target nuclei per kilogram, $\rho_\chi$ is the 
local WIMP energy density, $\sigma_n$ is the 
cross-section for scattering off a single nucleon, 
$m_N$ is the nucleon mass, $\mu_n$ is the reduced mass of the nucleon-WIMP
system, $E_R$ is the recoil energy, 
$\cos \gamma$ is angle between the velocity of the earth and the recoil 
velocity of the nucleon as seen the earth's rest frame, $F^2(E_R)$ is the helm 
form factor describing the loss of coherence of the nucleus at large momentum 
transfer \cite{Helm:1956zz}, 
$v_0 \simeq 220$~km/s is the WIMP velocity dispersion, 
$v_{\rm esc} \simeq 500$~km/s is 
the escape velocity, $\vec{v}_E$ is the velocity of the earth,
and $n(v_0,v_{\rm esc})$ normalizes the velocity distribution. 
We assume here a ``standard" Maxwellian distribution of WIMP velocities in the halo.
Conservation of energy and momentum dictate that the minimum velocity to scatter is,
\bea
v_{\rm min} = \sqrt{\frac{1}{2m_N E_R}} \left(\frac{m_N E_R}{\mu} +\delta 
\right).
\eea

In the case of fDM, the Wimp scattering is into one of the whole tower of 
excited states. For a given final state excited WIMP $j$ one has,
\bea
\kappa \to \kappa^j =  N_T \frac{\rho_\chi}{m_\chi} \frac{\sigma_n^j m_N}{2\mu_n}
\frac{(f_p Z + (A-Z)f_n)^2}{f_n^2},
\eea
and ,
\bea
v_{\rm min} \to v_{\rm min}^j = \sqrt{\frac{1}{2m_N E_R}} \left(\frac{m_N E_R}{
\mu} + \delta^j 
\right),
\eea
where as discussed above, we take $\delta^j \simeq j \delta$ and $\sigma^j_n \simeq \sigma_n$.  This results
in three relevant parameters controlling the predictions:
the WIMP mass $m_{\chi}$, the splitting between consecutive excited states $\delta$,
and the cross-section with nucleons $\sigma_n$.

\subsubsection{MfDM}

In MfDM scenarios the interactions between the WIMP and the target nucleus are 
mediated by photons which couple to the WIMP's magnetic dipole moment 
$\mu_\chi$.  The interaction with the target nucleus can either proceed through
its charge or magnetic moment, leading to both dipole - dipole (DD) and 
dipole-charge (DZ) interactions.  
The differential cross-section with respect to the recoil energy is,
\be
	\frac{d\sigma}{dE_R} = \sum_{i=1}^{N} ~  \frac{d{\sigma}_{DZ}^{i}}{dE_R}   + N  \frac{d{\sigma}_{DD}}{dE_R} 
	\label{dsigmadErMfDM}
\ee
\bea
	\frac{d{\sigma}_{DZ}^{i}}{dE_R} &=& \frac{4 \pi Z^2 \alpha^2}{E_R}  \left(\frac{{\mu}_{\chi}}{e}  \right)^2   \Big[   1  - \frac{E_R}{v^2}   \left( \frac{1}{2 m_N}  +   \frac{1}{m_\chi}  \right)   \nonumber \\ & ~~~ & -
	\frac{\delta^{i}}{v^2}  \left( \frac{1}{\mu_{N\chi}}  +  \frac{\delta^{i}}{2 m_N E_R}   \right)  \Big]   \left(\frac{S_{\chi}+1}{3 S_{\chi}} \right) F^2[E_R]
\label{DZMfDM}	
\eea
\bea	
	\frac{d{\sigma}_{DD}}{dE_R} = \frac{16 \pi {\alpha}^2 m_N}{v^2}   \left(\frac{{\mu}_{nuc}}{e} \right)^2   
	\left(\frac{{\mu}_{\chi}}{e}  \right)^2 \left(\frac{S_{\chi}+1}{3 S_{\chi}} \right) \left(\frac{S_N+1}{3 S_N}   \right) 
	{F_D}^2[E_R]
\label{DDMfDM}
\eea
where $N$ is the heaviest state kinematically accessible  \cite{Chang:2010en}.
A list of values for $Z$ and $\mu_n$ for a variety of relevant target nuclei is 
presented in Table~\ref{tab: isotopes}.
The DD term is proportional to $\mu_{nuc}^2$, implying that the rate at DAMA
(whose target is NaI) can be significantly enhanced as compared to
experiments using target nuclei with lower $\mu$, such as xenon or germanium.
	
In Eqs.~(\ref{dsigmadErMfDM})-(\ref{DDMfDM}), $F^2[E_R]$ is the 
Helm form factor \cite{Helm:1956zz} 
and ${F_D}^2[E_R]$ is the nuclear magnetic dipole
form-factor. While the helm form factor is expected to be
a good approximation for the factor by 
which the cross section decreases for 
non-zero momentum transfer in spin independent interactions, 
the nuclear magnetic 
dipole moment form-factor deserves more careful treatment. 
The usual thin-shell model for spin-dependent interactions need not be a 
good approximation for heavier nuclei~\cite{Engel:1992bf}. 
We follow Ref.~\cite{Toivanen:2009zz} in obtaining the form factors for 
${}^{129}$Xe, ${}^{131}$Xe, ${}^{133}$Cs and 
${}^{127}$I.
The theoretical model therein from which we have derived our form factor reproduces observables like
magnetic moments of relevant nuclear states quite well. Nonetheless,
as discussed in more detail in Ref. \cite{Chang:2010en}, this remains a source of uncertainty in our predictions which is
difficult to quantify in the absence of more direct experimental inputs.
We define the form factor as
\be
F^2(E_r)=\frac{1}{N} \Big( f_0^2 \Omega_0^2 F_{00}^2(Er) + 2 f_0 f_1 \Omega_0 \Omega_1 F_{01}(E_r) + f_1^2  \Omega_1^2 F_{11}^2(E_r)  \Big)
\ee
where $\Omega_0$, $\Omega_1$ are effective \textit{g} factors (Table IV, Ref.~\cite{Toivanen:2009zz}), and $F_{00}$, $F_{01}$,  $F_{11}$ are spin structure 
functions (Figs. 3-4, Ref.~\cite{Toivanen:2009zz}). The coefficients $f_0=\mu_n +\mu_p$ and $f_1=\mu_p-\mu_n$ are the isoscalar and isovector coupling constants. 
$\mu_n=-1.9 \mu_N$ and $\mu_p=2.8 \mu_N$ are magnetic moments of the proton and neutron.
The overall factor N is fixed by imposing the condition that the form factor is 1 at zero momentum transfer.

For all other nuclei we use the distribution~\cite{Lewin:1996} which takes into account the coupling to all `odd-group' nucleons ,
\be
 F^2(q r_n)  = \left\{ 
  \begin{array}{l l}
     j^2_0(q r_n) ~\quad (q r_n < 2.55, q r_n > 4.5)   \\
    0.047       \quad       \quad (2.55 <q r_n < 4.5)  \\
  \end{array} \right. ~.
  \ee

\begin{table}[t]
\begin{center}
\begin{tabular}{| c | c  c  c   c |}
\hline
Isotope     &   Z   &   Abundance(\%)  &  Spin   &   ${\mu}_{nuc}/{\mu}_N$   \\
                
\hline
\hline
${}^{17}$O  &   8       &     0.038      &      5/2    &     -1.894 \\
${}^{19}$F  &   9       &    100      &      1/2    &     2.629 \\
${}^{23}$Na  &   11       &    100      &      3/2    &     2.218 \\
${}^{43}$Ca  &   20       &    0.135      &      7/2    &     -1.317 \\
${}^{73}$Ge  &   32       &    7.76      &      9/2    &     -0.879 \\
${}^{127}$I  &   53       &    100      &      5/2    &     2.813 \\
${}^{129}$Xe  &   54       &    26.40      &      1/2    &     -0.778 \\
${}^{131}$Xe  &   54       &    21.23      &      3/2    &     +0.692 \\
${}^{133}$Cs  &   55       &    100      &      7/2    &     +2.582 \\
${}^{183}$W  &   74       &    14.31      &      1/2    &     +0.118 \\
\hline

\end{tabular}
\end{center}
\caption{\label{tab: isotopes} Natural isotopes with quantities relevant to direct 
detection searches~\cite{Lide:2010}.}
\end{table}

\subsection{Fits to the DAMA/LIBRA Signal}
\label{sec:scan}

Our next task is to assess the ability of fDM and MfDM to fit the DAMA 
observations.  We scan over $m_\chi$ and $\delta$, and construct a $\chi^2$
function to fit the amplitude over the $2\, -\, 8$~keVee energy 
bins reported by DAMA/LIBRA, 
using the quenching factors 0.3 (0.09) for Na (I)~\cite{Bernabei:1996vj}. 
To construct the $\chi^2$-function, we estimate the modulated amplitude as half
of the difference between the maximal and minimal scattering rates during the 
year\footnote{This approximation breaks down if the scattering rate were to be 
zero for a substantial part of the winter months. However we find that this
approximation is valid for all of the DAMA/LIBRA preferred regions of 
parameter space.}. We focus on the $2\, -\, 8$~keVee region, since for higher 
energies both the observed and predicted spectra are consistent with zero 
modulation and the inclusion of these energy bins would act to artificially 
improve the quality of the fit.

 \begin{figure}
\begin{minipage}[b]{0.5\linewidth}
\centering
\includegraphics[width = 3 in ]{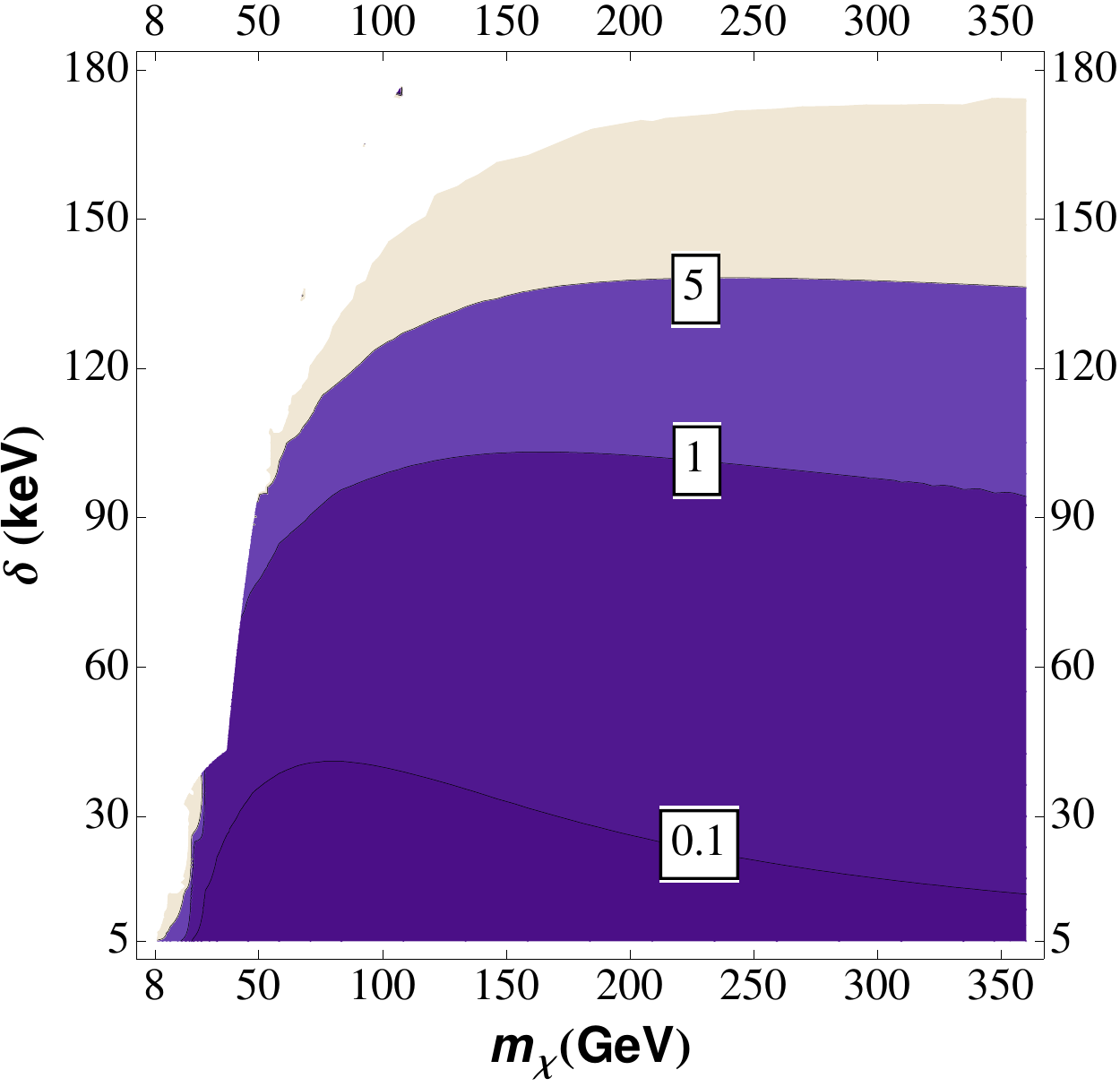} 
\end{minipage}
\hspace{-0.5 cm}
\begin{minipage}[b]{0.5\linewidth}
\centering
\includegraphics[width = 3 in]{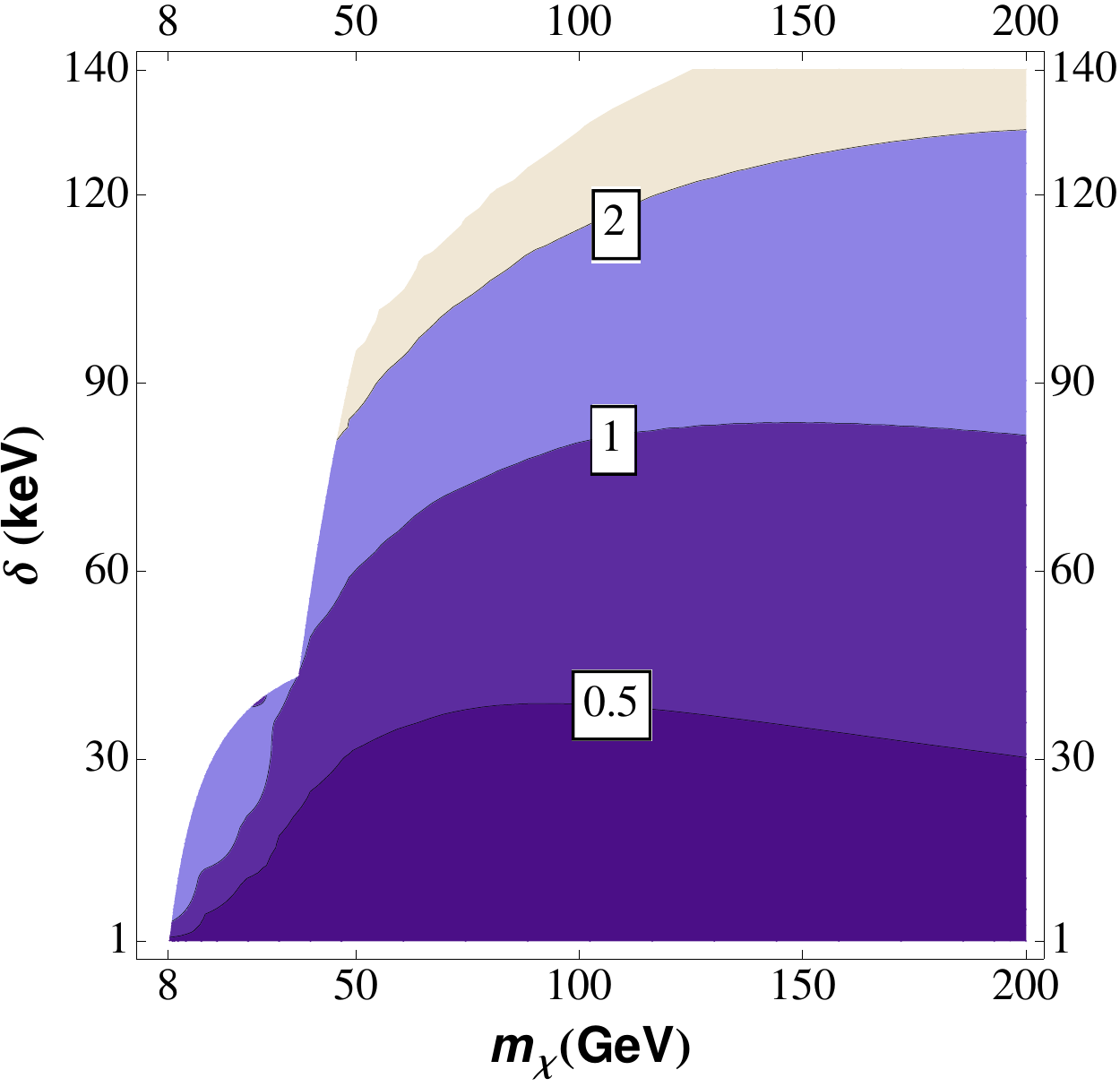} 
\end{minipage}
\caption{\label{fig:fDMMfDMscales} Best fit values
of ${\sigma}_n^0$ in units of $10^{-40}$ cm$^2$ (left panel) in the fDM scenario
and ${\mu_{\chi}}$ in units of $10^{-3} \mu_N$
(right panel) in the MfDM scenario.
The maximum value in both cases is 
along the edge of the tan (off-white) region and is 24 in the fDM case and  2.8 in 
the fDM case. The white region in both plots is not allowed as the $\delta$ is too high for any scattering to take place.
 The kink in the boundary of the allowed region at $m_{\chi} \sim40$ 
GeV is due to the fact that below this mass the upper limit on $\delta$ is set by scattering off of Sodium, and 
above it the limit is set by scattering off of Iodine. 
}
 \end{figure}

For each point in the $m_\chi-\delta$ plane, we find the corresponding
spin-independent scattering cross-section $\sigma_n$ for fDM or the
dipole moment $\mu_\chi$ for MfDM that leads to the minimal value of $\chi^2$.
The results are shown in Fig. \ref{fig:fDMMfDMscales}.
Using these values of $\sigma_n$ for fDM and $\mu_\chi$ for MfDM we can 
calculate the scattering rates at a particular direct detection experiment 
using Eq.(\ref{dRdErdcosgamma}) and Eq.(\ref{dsigmadErMfDM}) respectively.
In this sense, the bounds on regions of $m_\chi$ and $\delta$
that we later derive from other experiments 
are themselves
dependent on DAMA.   We consider 
bounds from XENON100 (both the 11.17 days of 
data collected in 2009~\cite{Aprile:2010um} and the 100 days of data collected 
in 2010~\cite{Aprile:2011hi}); CDMS II (including the low energy threshold 
analysis)~\cite{Ahmed:2009zw}; KIMS~\cite{Lee.:2007qn}; and 
COUPP~\cite{Behnke:2010xt}. Bounds from XENON10~\cite{Angle:2009xb}, 
ZEPLIN III~\cite{Akimov:2010vk}, 
CRESST II~\cite{Angloher:2008jj,Angloher:2011uu}, and 
PICASSO~\cite{Archambault:2009sm} were considered, but do not appear in the 
figures as they are weaker than the bounds from other experiments.
A summary of the data used to constrain fDM and MfDM
scenarios is presented in Table~\ref{tab:DM_prof}.

\begin{figure}
\begin{minipage}[b]{0.5\linewidth}
\centering
\includegraphics[width = 3 in ]{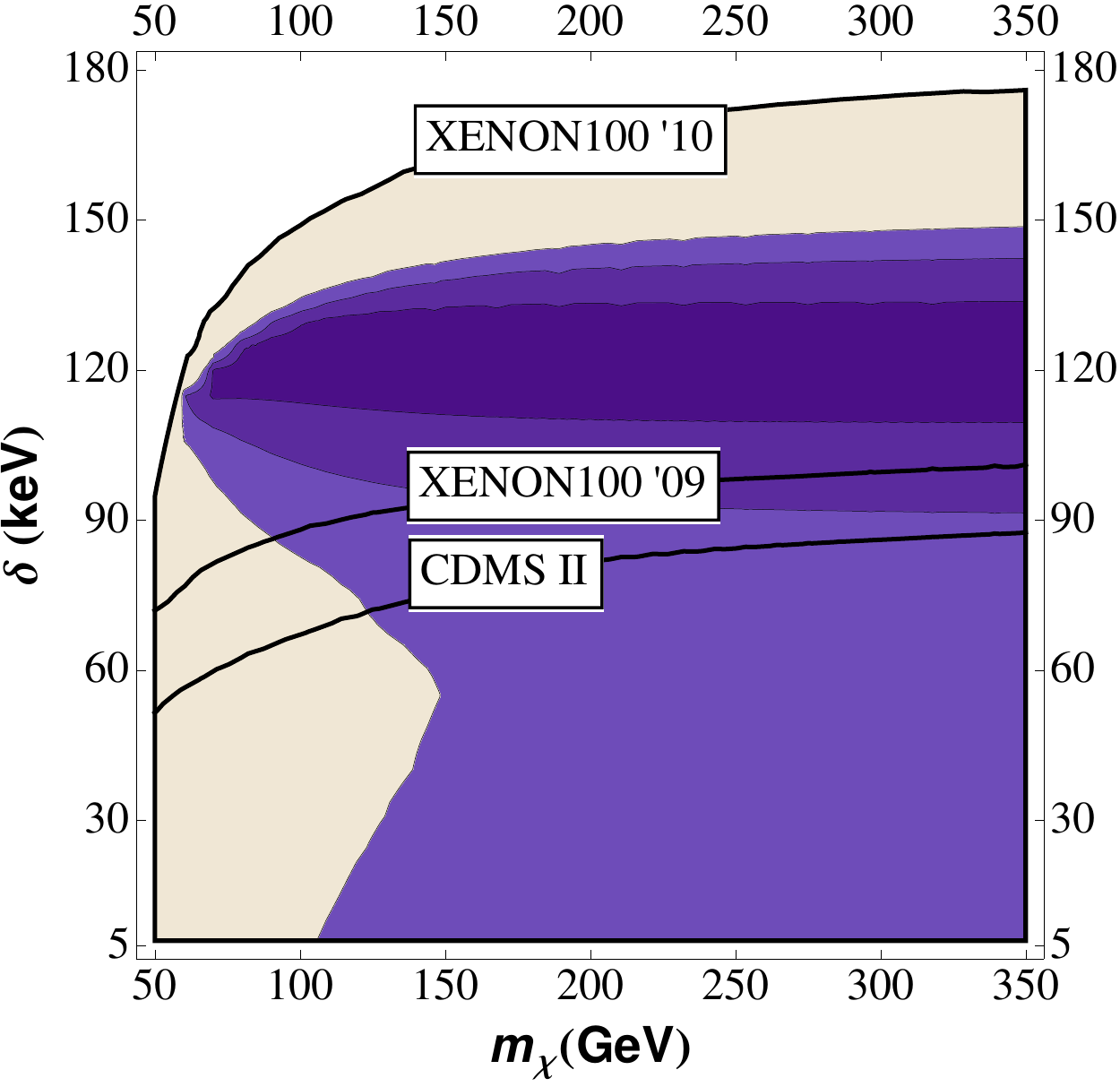} 
\end{minipage}
\hspace{-0.5 cm}
\begin{minipage}[b]{0.5\linewidth}
\centering
\includegraphics[width = 3 in]{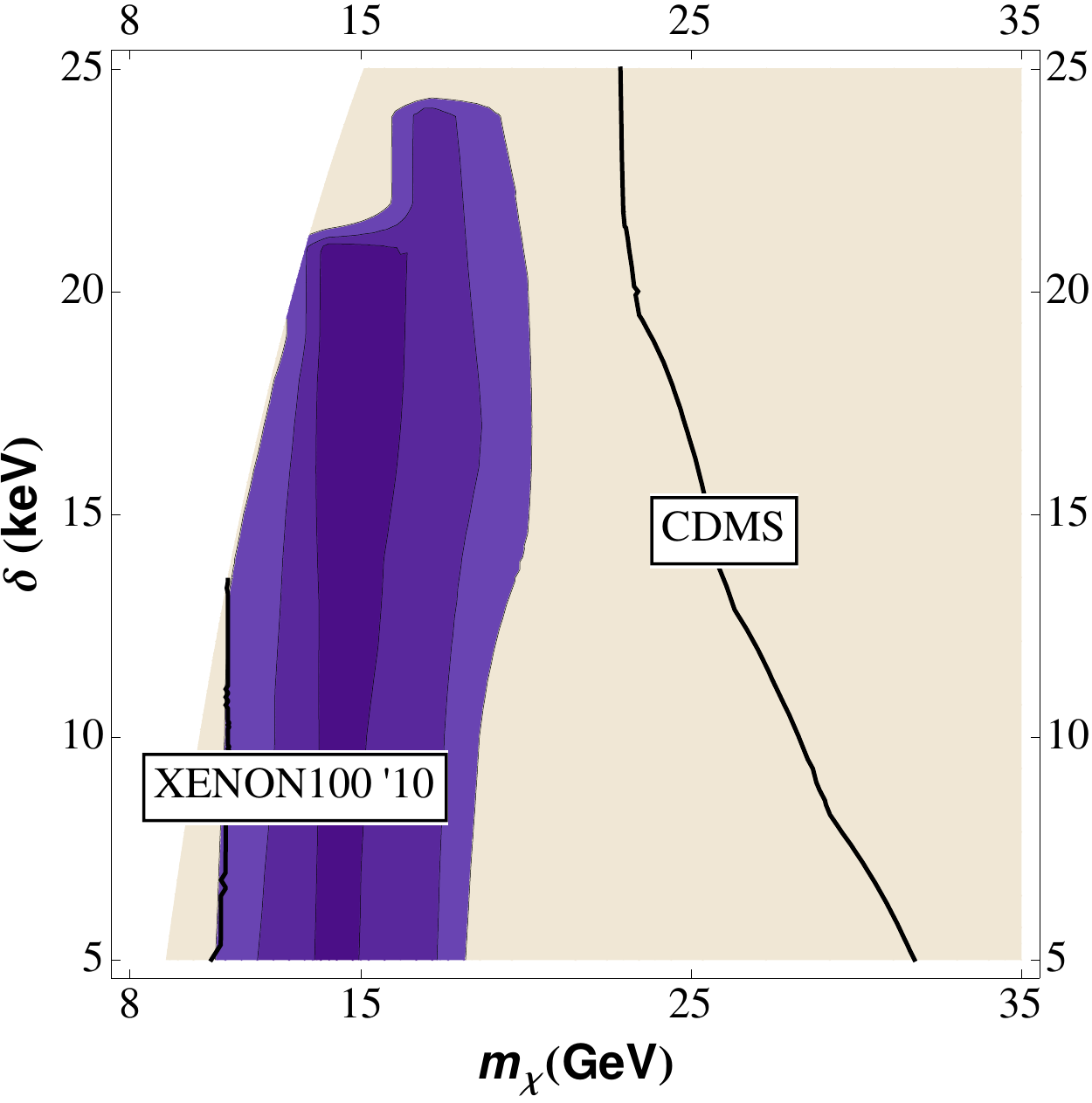} 
\end{minipage}
\caption{\label{fig:fDMplots} Constraints on fDM, for large $(m_\chi,\delta)$ 
(left) and small $(m_\chi,\delta)$ (right), from direct detection experiments. 
The darker blue (grey), blue (grey) and lighter blue (grey) regions correspond to
the fDM points that fit the DAMA/LIBRA annual modulation spectrum at $68\%$, 
$95\%$, and $99.7\%$ C.L. In the left panel regions to the right and below
of each solid black line are ruled out by the corresponding direct detection 
experiment. In the panel on the right, parameter space to the left 
of the line corresponding to the CDMS low threshold limit 
and to the right of the line from XENON10 '10 are excluded.
}
 \end{figure}

\begin{figure}
\begin{minipage}[b]{0.5\linewidth}
\centering
\includegraphics[width = 3 in ]{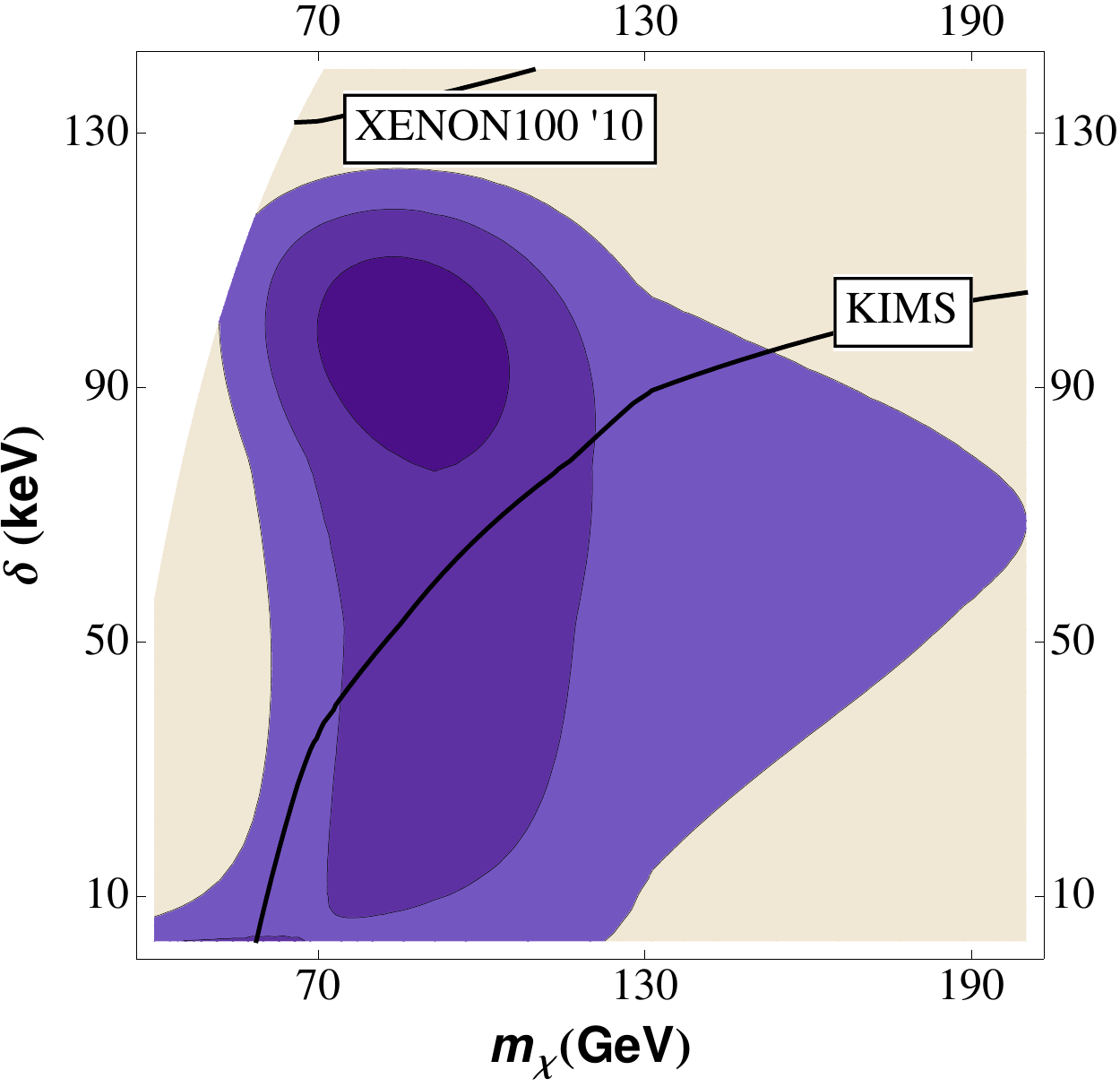} 
\end{minipage}
\hspace{-0.5 cm}
\begin{minipage}[b]{0.5\linewidth}
\centering
\includegraphics[width = 2.8 in]{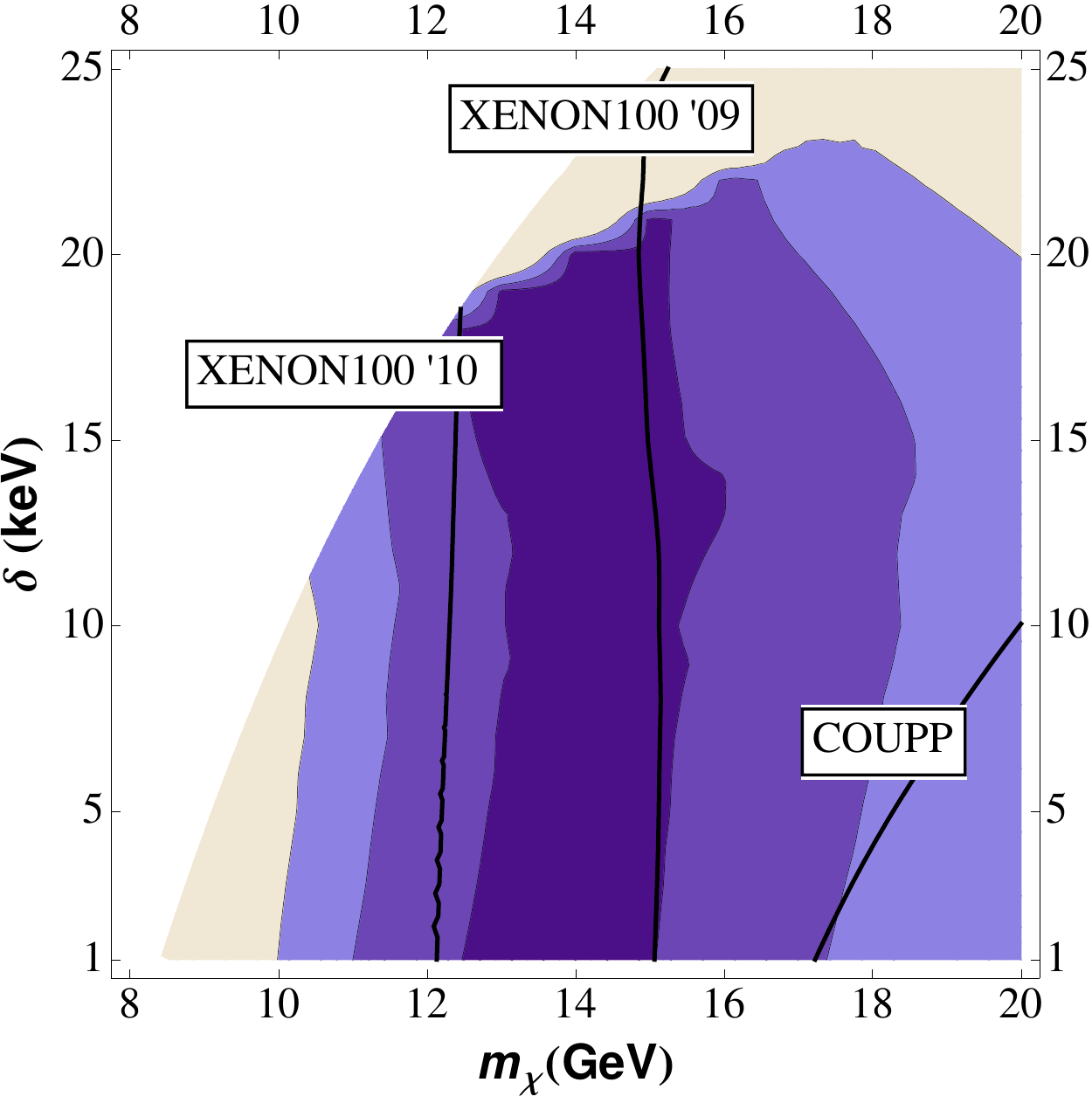} 
\end{minipage}
\caption{\label{fig:MfDMplots} Constraints on MfDM, for large $(m_\chi,\delta)$ 
(left) and small $(m_\chi,\delta)$ (right), from direct detection experiments. 
The darker blue (grey), blue (grey) and lighter blue (grey) regions correspond to
the MfDM points that fit the DAMA/LIBRA annual modulation spectrum at $68\%$, 
$95\%$, and $99.7\%$ C.L. The parameter space below and to the 
right of each of the 
black lines are ruled out by the corresponding experiment.}
\end{figure}

In the plane of $m_\chi$ and $\delta$, we find islands where 
both fDM and MfDM can provide an acceptable fit  within error bars to the 
DAMA/LIBRA data.  We map out the contours of $68\%$, $95\%$, and $99.7\%$ C.L. 
agreement with the DAMA/LIBRA annual modulation amplitude, which we 
plot as the darkest blue (grey), dark blue (grey), and lighter blue (grey) 
shaded regions of Fig.~\ref{fig:fDMplots} and Fig.~\ref{fig:MfDMplots} for both fDM 
and MfDM, and for low mass and high mass WIMPs.  
These contours are somewhat conservative in the sense
that the best fit points have ${\chi}^2 <1$ per degree of freedom.
The tan (off-white) regions of
each plot have poorer agreement, and the white regions of
each figure have $\delta$ large enough that $v_{\rm min} > v_{\rm esc}$,
so no scattering is possible.

%

\begin{table}[t]
\begin{center}
\hspace*{-0.5cm}
\begin{tabular}{|l||c|l|c|c|c|}
\hline
~~~~~Experiment  &  Exposure    &   ~~Time~Period             &    Signal Window  &   Observed   \\
\hline
\hline
CDMS '08     &  194.1 kgd   & 7/1/07 - 9/1/08     &      10 - 100 keV     &             2                    \\
CDMS Low-Energy & 241 kgd       &      101/06 - 9/1/08         &        2 - 5 keV          &           324             \\
XENON10    &     $0.3\times316.4$ kgd    &  10/6/06 - 2/14/07      &  4.5 - 75 keV     &     13            \\
XENON100    &    161 kgd             &    10/20/09 - 9/12/09  &    7.4 - 29.1 keV  &     0             \\
XENON100   &   $48\times100.9$ kgd    &    1/13/10 - 6/8/10     &     8.4 - 44.6 keV    &     3   \\
ZEPLIN III      &   $0.5\times63.3$ kgd    &    2/27/08 - 5/20/08   &   17.5 - 78.8 keV     &     5         \\
CRESST II (W)      &   $0.59\times0.9\times48$  kgd     &    3/27/07  - 7/23/07   &     12 - 100 keV           &     7       \\
CRESST(O)    &      564 kgd                 &   7/09 - 10/10       & $\sim$10 - 40 keV      &    32           \\
CRESST(Ca,W,O) &  730 kgd     &   7/09 - 3/11     & $\sim$10 - 40 keV    &    67        \\
KIMS                 &       3409 kgd     &     1/05 - 12/06          &    20 - 100 keV       &       955  \\
COUPP    &  28.1 kgd  &    11/19/09 - 12/18/09 &   21- 200 keV   &   3        \\
PICASSO   &   13.75 kgd  &  6/07 - 7/08    &  20 - 200 keV    & 2051   \\
\hline
\end{tabular}
\caption{\label{tab:DM_prof} Exposure and data collection period, signal energy window, and 
number of observed events at various dark matter direct detection experiments.
 For CRESST we write $\sim 10$ keV
for the sake of brevity in the table. The threshold is different for each of their detector modules. 
The Observed events at KIMS has been 
inferred using the exposure and event rates from Ref.~\cite{Lee.:2007qn}.  }
\end{center}
\end{table}

\subsubsection{fDM}

In the case of fDM, the high WIMP mass region is shown in the left panel of 
Fig.~\ref{fig:fDMplots}.  The regions that fit the DAMA/LIBRA data well are 
approximately the same as the those previously identified in 
Ref.~\cite{TuckerSmith:2004jv,Chang:2008gd} as being preferred by
iDM.  As such, this parameter space represents a WIMP
which is not particularly fluffy because 
only one excited state is relevant for the scattering.  It thus
suffers the same fate as standard iDM and is highly constrained by XENON100
and CDMS for a Maxwellian halo \cite{Farina:2011bh}.

In the low mass region, there is a separate island of good fit region, where the
scattering is dominantly off of Na nuclei.  Here, we find roughly
any $\delta \lesssim 20$ keV works to describe DAMA, which for small splitting
looks very much like a fluffy WIMP.  However,
WIMP masses above around 11 GeV
are excluded by XENON 100 (which is essentially enough to remove the
interesting parameter space) and the CDMS low threshold analysis removes the
entire interesting low mass region as well.
It is interesting to note that unlike the other exclusion limits, 
the low threshold CDMS limit excludes the region to its left, for smaller $m_\chi$.
This is due to the fact that the values of $\sigma_n$
extracted from fitting to DAMA/LIBRA increase as $m_\chi$ decreases, 
as can be seen in 
the left panel of Fig.~\ref{fig:fDMMfDMscales}, which
compensates the smaller typical scattering energy all the way to the region
where no scattering is possible.

Combining the low and high mass regions, we see that the regions of fDM
parameter space consistent with the DAMA energy spectrum are strongly 
constrained by the null search limits from other direct detection 
experiments.

\subsubsection{MfDM}

The results for MfDM in the high WIMP mass range ($m_\chi \sim 100$~GeV)
are shown in the left panel in Fig \ref{fig:MfDMplots}.  The DAMA/LIBRA 
spectrum is well fit in the region $m_\chi \sim$~80 GeV and $\delta \sim 100
$~keV. The strongest constraints obtained are from XENON100 and KIMS 
(whose Cs 
and I nuclei have relatively high nuclear magnetic moments, and are thus 
particularly good probes of MfDM). XENON100 sets the stronger constraint 
and excludes the entire region allowed by DAMA. Similarly to the fDM scenario,
this region of parameter space does not correspond to a very fluffy WIMP
and the region of parameter space favored by DAMA/LIBRA is similar to that
found in the MiDM scenario of Ref.~\cite{Chang:2010en}.

In the low WIMP mass region, we find that masses around 14~GeV and with
$\delta \lesssim 20$~keV provide a good fit to DAMA.  
This region is shown in the right panel of 
Figure~\ref{fig:MfDMplots}, where we see relevant constraints from
XENON100 and COUPP.  CDMS low threshold constraint is weak, due to the 
small magnetic coupling,  while the XENON100 limit only constrains a part of 
the parameter space that provides a good fit to the DAMA/LIBRA data.
The XENON100 limit allows a significant portion of the MfDM parameter 
space that fits the DAMA/LIBRA signal at the $> 95\%$ C.L.
 The fit in this allowed
region is also not dependent on $\delta$ for values less than $15$ keV.

The larger $\delta$ region indicates that MiDM with WIMP masses around 11 GeV
is consistent with current direct detection data. 
The small $\delta$ region
realizes the hope of a fluffy WIMP, with many excited states contributing to the
scattering.  
For example, an 11 GeV WIMP can scatter up to states that are heavier by 
$\lesssim 15$ keV, so for a splitting of $\delta=1$ keV, 15 excited
states participate in the scattering.  For smaller values of $\delta$, the
number of relevant states in the scattering increases.

\section{Outlook}
\label{sec:concl}

We have considered a picture where dark matter is the lowest lying state in a
composite sector, whose low energy scattering with nuclei is inelastic.  
We refer to such a WIMP as fluffy, since its internal structure as a confined state
is rather easily perturbed by its environment.
For very low 
confinement scales, the preferred
inelastic splitting emerges somewhat organically, driven
largely by the energy thresholds of DAMA itself.  We perform fits and
find that light ($\sim 11$~GeV) WIMPs with small splittings $\delta$ provide the
best fits to the DAMA observation of annual modulation and energy spectrum.
For a WIMP whose interactions
with the SM are through iso-spin conserving spin-independent couplings to the
SM, constraints (largely from XENON 100) are enough to close off the region
of parameter space able to explain the DAMA signal.  However, a fluffy WIMP
whose interactions are magnetic in nature can explain DAMA and still remain
consistent with the bounds from other direct detection experiments.   Because of
the low masses favored by the fit, it is difficult for XENON to access the
parameter space of interest.  Lower threshold experiments perhaps offer the most
promising probes in the future.

A low compositeness scale is a challenge for cosmology, and leads to potentially
many unusual features for a theory of dark 
matter.  To begin with,
our analysis assumed the bulk of the
WIMPs in the halo were in the ground state.  If this were not the case, the
WIMPs would down-scatter as well as up-scattering.  One could imagine
engineering this possibility directly
into a model of luminous dark matter \cite{Feldstein:2010su}, which is
an interesting independent line of investigation beyond the scope of this work,
but we prefer to imagine there are interactions which efficiently de-excite
the WIMPs, perhaps through a coupling to neutrinos \cite{Falkowski:2009yz}.

In the early Universe, a fluffy WIMP can undergo a freeze-out process which
differs substantially from a standard 
WIMP. The seed partons may freeze out at early times, but the confined states
may come back into equilibrium because of the 
surrounding clouds of light partons increase
their effective cross sections after the phase transition.  
This can lead to an additional dilution of the
WIMPs at late times
\cite{Kang:2006yd,Alves:2010dd,Feng:2011ik}.  Alternately, one could explore
the case where excitations with large splittings continue to play an active
cosmological role, perhaps obviating the need for an external stabilization
symmetry \cite{Dienes:2011ja,Dienes:2011sa}.

Since we favor a low confinement scale, there will be additional (at least) gauge
degrees of freedom contributing to the thermal bath even at relatively late times.
Precision measurements of the primordial abundances of light elements produced
through big bang nucleosynthesis, and of the cosmic microwave background,
provide tight constraints on additional light degrees of freedom during the relevant
epochs \cite{Fields:2006ga,Komatsu:2010fb}.  Both sets of measurements
are currently
consistent with no new degrees of freedom being present, with error bars
allowing for around at most one new state equivalent to an``effective neutrino
species".
Future CMB measurements by PLANCK are expected to have the precision
to shrink the uncertainties to the order of a tenth of an effective neutrino species
\cite{Hamann:2007sb}.  One could imagine evading these bounds if the
dark sector has a separate temperature from the
SM plasma \cite{Feng:2011ik}, or if the coupling of
the new confined force is time-dependent, perhaps being set by value of a
modulus through a term such as $\phi F^{\mu \nu} F_{\mu \nu}$, where
$\langle \phi \rangle$ starts at large values, leading to a tightly bound WIMP
during the cosmologically sensitive times, but rolls to lower values in late
cosmological times, leading to the WIMPs puffing up.

After confinement, one can expect the
analogues of glueball states for the new 
gauge force (whose masses should be roughly $\Lambda \sim$ keV) could
contribute to relevant phenomena.  
For example, if sufficiently long-lived, they could end up as 
a subdominant component
of warm dark matter, or their decays could contribute additional entropy
to the Universe.
In addition, 
such states may have large (strong force residual) interactions with the WIMPs.
If WIMPs can efficiently lose 
kinetic energy in collisions, either by converting kinetic energy to excitation
energy and then de-exciting by radiating a glueball, or by exchanging glueballs in 
elastic collisions, it can cause elliptical galaxies to become spherical
\cite{Feng:2009hw,Buckley:2009in}.  One can even
imagine more radical shifts in galactic dynamics, such as cases where some
fraction of the binding is due to glueball exchange (in addition to gravity), with the
galaxy itself looking something like the analogue of a heavy nucleus state of
the new gauge force.

One can imagine other scenarios in which a fluffy WIMP might provide an 
interesting model of dark matter.  For example, models of
exciting dark matter \cite{Finkbeiner:2007kk} invoke an
MeV split excited WIMP to explain the INTEGRAL/SPI 511 keV gamma ray
excess \cite{Weidenspointner:2006nua}.  
One could easily imagine a fluffy WIMP model allowing this
small scale to emerge organically as it did here in an inelastic scattering context.
It would be interesting to see if a common framework could potentially explain the
DAMA signal as well as the INTEGRAL excess within a common framework
relying on a single value of $\delta$.

These potential features 
are interesting, and highlight both the challenges in designing
a workable cosmology, as well as motivating explorations of some truly novel
phenomena.  While straw man constructions are easy to construct piece by piece,
a compelling, unified framework would be worth pursuing.  We leave a
detailed exploration of these ideas for future work.

\section*{Acknowledgments}
\label{sec:ack}
We are grateful for helpful conversations with Jonathan Feng and James Bjorken.  TT 
is pleased to acknowledge the SLAC theory group for their hospitality during his 
many visits, and to the KITP (supported in part by the NSF under Grant No. 
PHY05-51164) where part of it was performed.  The work of TT is supported in part by
the NSF under grant PHY-0970171. AM was supported at IIT by DOE grant number
DE-FG02-94ER40840 and at University of Oregon by DOE grant number DE-FG02-96ER40969.

\end{document}